\documentclass[a4paper,12pt]{article}
\usepackage{amsmath,amsfonts,amssymb}
\usepackage{graphicx}

\newcommand{\beq}{\begin{equation}}
\newcommand{\eeq}{\end{equation}}
\newcommand{\bea}{\begin{eqnarray}}
\newcommand{\eea}{\end{eqnarray}}
\newcommand{\ba}{\begin{array}}
\newcommand{\ea}{\end{array}}
\newcommand{\bec}{\begin{center}}
\newcommand{\eec}{\end{center}}
\newcommand{\bei}{\begin{itemize}}
\newcommand{\eei}{\end{itemize}}

\newcommand{\TeV}{\,\mathrm{TeV}}
\newcommand{\GeV}{\,\mathrm{GeV}}

\def\unit{\relax{\rm 1\kern-.26em I}}

\title{}
\begin{document}

\thispagestyle{empty}

\vspace{-2cm}

\begin{flushright}
{\small CPHT-RR102.1110 \\
LPT-ORSAY 10-87 \\
SACLAY-T10/175} \\
\end{flushright}
\vspace{1cm}

\begin{center}
{\bf\LARGE  On messengers and metastability\\ in gauge mediation}

\vspace{0.5cm}

{\bf Emilian Dudas}$^{a,b}$, {\bf St\'ephane  Lavignac}$^{c}$
{\bf and 
Jeanne Parmentier}$^{a,c}$
 \vspace{0.5cm}

{\it $^{a}$ Centre de Physique Th\'eorique\! \footnote{Unit\'e mixte
du CNRS (UMR 7644).},
Ecole Polytechnique and CNRS,\\
F-91128 Palaiseau, France.\\[10pt]
$^{b}$ Laboratoire de Physique Th\'eorique\! \footnote{Unit\'e mixte du CNRS (UMR 8627).},
Universit\'e de Paris-Sud,\\
B\^at. 210, F-91405 Orsay, France.\\ [10pt]
$^{c}$ Institut de Physique Th\'eorique\! \footnote{Laboratoire
de la Direction des Sciences de la Mati\`ere du Commissariat \`a l'Energie
Atomique et Unit\'e de Recherche associ\'ee au CNRS (URA 2306).},
CEA-Saclay,\\
F-91191 Gif-sur-Yvette Cedex, France.
}

{\bf Abstract}
\end{center}

One notoriously difficult problem in perturbative gauge mediation
of supersymmetry breaking via messenger fields is the generic presence of
a phenomenologically unacceptable vacuum with messenger vevs, with
a lower energy than the desired (``MSSM'') vacuum.
We investigate the possibility that quantum corrections promote the latter
to the ground state of the theory, and find that this is indeed feasible.
For this to happen, the couplings of the messengers to the goldstino
superfield must be small, and this implies an additional suppression
of the MSSM soft terms with respect to the supersymmetry breaking scale.
This in  turn sets a lower limit on the masses
of the messengers and of the supersymmetry breaking fields,
which makes both sectors inaccessible at colliders.
Contrary to other scenarios like direct gauge mediation, gaugino
masses are unsuppressed with respect to scalar masses.

\clearpage

\begin{flushright}
{\today}
\end{flushright}

\tableofcontents

\section{Introduction and conclusions}
\label{sec:intro}

Gauge mediation of supersymmetry breaking~\cite{gr} is an attractive way
of solving the flavour problem of supersymmetric theories. In its minimal
version, it leads to a highly predictive spectrum which has been extensively
studied\footnote{More recently, there has been an intense activity
aiming at providing generalized gauge mediation models~\cite{cfs,dlp,dkk} and at studying their phenomenology \cite{abel}.}
from a phenomenological viewpoint~\cite{gaugepheno,Drees_book}.
On the other hand, the construction of an explicit supersymmetry breaking
sector\footnote{For a review of the recent progress on the subject, see
e.g. Ref.~\cite{is}.} coupled to messenger fields~\cite{gauge2}
responsible for the generation of the MSSM soft terms leads to instabilities
of the scalar potential in the messenger direction, and therefore to
dangerous vacua breaking the electric charge and colour. While the
desired MSSM vacuum can be locally stable with a lifetime exceeding
the age of the Universe~\cite{ISS06}, it would clearly be more satisfactory
to avoid the messenger instabilities. Recently progress was made in this
direction in scenarios in which messengers are part of the supersymmetry
breaking sector, dubbed direct gauge mediation models~\cite{direct}.
However these generally have difficulties in generating
large enough gaugino masses, and more work is needed in order to
construct fully realistic models.

The purpose of  the present letter is to investigate whether it is
possible at all to avoid messenger instabilities in explicit, perturbative
supersymmetry breaking models coupled to messenger fields. Based
on the analysis of a specific class of models of the O'Raifeartaigh type,
we find evidence
that this is indeed possible, provided that the coupling of the messengers
to the goldstino superfield is sufficiently suppressed with respect to
their couplings to other fields from the supersymmetry breaking
sector. We shall consider the following class of models, written below
in the canonical form of Refs.~\cite{ray,ks}:
\beq
W  \ = \ f \, X + \, \frac{1}{2} \left( h_a^{(X)} X  +  h_a^{(\chi_i)} \chi_i \right) \varphi_a^2\,
  +\,  m_a \varphi_a Y_a\,  +\,
  \phi \left( \lambda_X X + \lambda_i \chi_i + M \right) \tilde \phi \ ,
\label{i1}
\eeq
where $X$ is the goldstino superfield, $\chi_i$, $\varphi_a$ and
$Y_a$ are the other fields needed to break supersymmetry, and
($\phi$, $\tilde \phi$) are the messenger superfields. Here and
in the following, summation over repeated indices is understood.
Notice that the R-symmetry of the O'Raifeartaigh sector~\cite{ns}
is broken by the messenger couplings.
As we are going to show in Section~\ref{sec:oneloop}, a necessary
condition for avoiding messenger instabilities in the one-loop
effective potential reads (written for simplicity in the case of equal
O'Raifeartaigh masses $m_a = m$ and with all couplings evaluated
at the scale $\mu = m$):
\beq
|\lambda_X| \  < \ \frac{1}{8 \pi^2} \ | \sum_a h_a^{(X)} (\lambda\! \cdot\!{\bar h}_a) | \ ,
\label{i2}
\eeq
where $(\lambda\! \cdot\!{\bar h}_a) \equiv \sum_i \lambda_i {\bar h}_a^{(\chi_i)}$.
This result is valid when the masses of the O'Raifeartaigh fields
are small compared with the messenger mass $M$.

We emphasize that, once the condition~(\ref{i2}) is imposed,
gaugino and scalar masses of the same order of magnitude are
generated by loops of messenger fields. In particular, there is no
contradiction between the one-loop stability of the MSSM vacuum
and non-vanishing gaugino masses. This is to be contrasted with
the tree-level supersymmetry breaking models discussed in
Ref.~\cite{ks}, in which gaugino masses are not generated at the
one-loop level and at leading order in supersymmetry breaking.
The class of models we consider evade the conclusions of
Ref.~\cite{ks} because the pseudo-modulus space is not stable
at $\lambda_X X + \lambda_i \chi_i + M = 0$.

As we are going to see in Section~\ref{sec:oneloop}, the
one-loop stability of the MSSM vacuum requires heavy messenger
and O'Raifeartaigh fields, which are therefore out of reach of the LHC.

\section{Generic O'Raifeartaigh models coupled to messenger fields}
\label{sec:framework}

In this section, we review the tree-level vacuum structure of generic
O'Raifeartaigh models coupled to messenger fields, and point
out the instability of the scalar potential in the messenger direction.
We adopt the parametrization of Refs.~\cite{ks,iss2}:
\beq
W \ = \ X_i \, f_i (\varphi_a)\, +\, g(\varphi_a) \, + \, \phi
\left( \lambda\! \cdot\! X  +M \right) \tilde{\phi} \ ,
\label{gf1}
\eeq
where $X_i$, $i = 1 \cdots N$ and $\varphi_a$, $a=1 \cdots P$
are O'Raifeartaigh fields, ($\phi$, $\tilde \phi$) are messenger fields,
and we have defined $\lambda\! \cdot\! X \equiv \sum_i \lambda_i X_i$.

\subsection{Tree-level vacuum structure and messenger instability}

The F-term equations of motion are given by:
\bea
&& - {\bar F}_i = f_i (\varphi_a) + \lambda_i \phi \tilde{\phi} \ , \quad
- {\bar F}_a
=  X_i \ \partial_a f_i (\varphi_b) + \partial_a g(\varphi_b) \ ,
\nonumber \\
&& - {\bar F}_{\phi} \ = \ (\lambda\! \cdot\! X  +M ) \, \tilde{\phi} \ , \quad
- {\bar F}_{\tilde \phi} \ = \ \phi \, (\lambda\! \cdot\! X  +M ) \ ,
\label{gf2}
\eea
and the tree-level scalar potential reads:
\bea
&& V \ = \ \sum_i \left|f_i (\varphi_a) \ + \ \lambda_i \phi
\tilde{\phi} \right|^2 \, + \ \sum_a \left| X_i \partial_a f_i(\varphi_b)
+ \partial_a g(\varphi_b)  \right|^2 \nonumber \\
&& + \ |\lambda\! \cdot\! X  + M|^2 \, (|\phi|^2 + |{\tilde \phi}|^2) \ .
\label{gf4}
\eea
In the following, we will assume that the MSSM D-term vanish
at the minimum of the scalar potential, such that
$\langle \phi \rangle = \langle \tilde \phi \rangle$. Let us now
review the conditions for tree-level supersymmetry breaking,
assuming generic functions $f_i$ and $g$.
Supersymmetry is broken for $N > P$ in the absence of
messenger fields, and for $N > P+1$ when they are present.
With this condition, the equations
$F_a= F_{\phi}=F_{\tilde \phi}= 0$ can always be
satisfied, leaving $N-P$ tree-level flat directions, which are linear
combinations of the fields $X_i$.  The vevs of the fields $\varphi_a$,
on the contrary, are completely determined and the functions
$f_i(\varphi_a)$ can be chosen such that
$\langle \varphi_a \rangle = 0$ (this will be the case of the
models we will specialize to in Section~\ref{sec:oneloop}).

The models defined above possess two tree-level vacua
(or more precisely two local minima extending to pseudo-moduli
spaces):
\begin{itemize}
\item a vacuum with vanishing messenger vevs,
$\phi =\tilde{\phi}=0$, and energy
\beq
V_1 \ = \ f^2 \, ,
\label{gf5}
\eeq
where we have defined $f^2 \equiv \sum_i {\bar f}_i f_i$.
This is the phenomenologically desired vacuum, and we shall
refer to it as the MSSM vacuum.
\item a vacuum with non-vanishing messenger vevs
\beq
\phi \tilde{\phi} \ = \ -\, \frac{1}{|\lambda|^2}\, \bar \lambda\! \cdot\! f \ ,
\label{gf6}
\eeq
located at $\lambda\! \cdot\! X  +M = 0$, and energy
\beq
V_2\ =\ \sum_i\, \left|f_i - \frac{\lambda_i}{|\lambda|^2}\,
{\bar \lambda}\! \cdot\!f \right|^2\,
=\ f^2 \ - \ \frac{| \bar \lambda\! \cdot\! f|^2}{|\lambda|^2} \ ,
\label{gf7}
\eeq
where we have defined
$|\lambda|^2 \equiv \sum_i {\bar \lambda}_i \lambda_i$
and $\bar \lambda\! \cdot\! f \equiv \sum_i {\bar \lambda}_i f_i$.
We shall refer to this vacuum as the messenger vacuum.
\end{itemize}
Comparing Eq.~(\ref{gf5}) with Eq.~(\ref{gf7}), we can see that the
unwanted messenger vacuum is the ground state of the model.
Moreover, the pseudo-moduli space extending the MSSM vacuum
is not stable everywhere: at $\lambda\! \cdot\! X  +M = 0$,
$\phi =\tilde{\phi}=0$ becomes a local maximum and one is driven
to the messenger vacuum. This is the vacuum stability problem
mentioned in the introduction.
The purpose of the present letter is to find appropriate conditions
ensuring that the MSSM vacuum is the global minimum of the one-loop
effective potential.

\subsection{Flat directions and their lifting}

In order to remove the instabilities of the tree-level vacuum along
the pseudo-moduli space, quantum corrections should stabilize
all flat directions of the O'Raifeartaigh sector\footnote{Notice that
if we accept to live in a metastable vacuum, this condition is actually
sufficient for phenomenological viability.}. A necessary condition
for this to happen is that the  $N-P$ flat directions appear in the
fermionic and/or scalar mass matrix, since these matrices determine
the one-loop effective potential through the Coleman-Weinberg
formula~\cite{cw}. In the absence of messenger fields, the fermionic
mass matrix takes the form:
\begin{eqnarray}
M_F\ =\
\begin{pmatrix}
 0  & \partial_a f_i (\varphi)  \\
  \partial_b f_j (\varphi)   & X_i \partial_a \partial_b f_i (\varphi)
  + \partial_a \partial_b g (\varphi)
\end{pmatrix} \ ,
\label{gf9}
\end{eqnarray}
in which the tree-level flat directions appear through the $P (P+1)/2$
combinations:
\beq
\chi_{ab} \ \equiv \ X_i \, \partial_a \partial_b f_i (\varphi) \ ,
\label{gf10}
\eeq
out of which ${\rm min} \{N, P(P+1)/2\}$ are independent.
It is easy to check that the scalar mass matrix depends on the same
combinations of fields. Taking into account the condition for
supersymmetry breaking, we arrive at the following necessary
conditions:
\beq
P + 1 \ < \ N\ \leq \ \frac{P (P+3)}{2} \ .
\label{gf11}
\eeq
%

\subsection{Comparison of the tree-level vacuum energies }

The purpose of this letter is to show that quantum corrections can
promote the MSSM vacuum to the ground state of the theory
at the price of suppressing the coupling of the messengers to
the low-energy goldstino superfield. For this to be possible,
the difference between the tree-level MSSM and messenger
vacuum energies,
\begin{eqnarray}
\Delta V \ \equiv\ V_1 - V_2\ = \ \frac{| \bar \lambda\! \cdot\! f|^2}{|\lambda|^2}\ ,
\label{gf8}
\end{eqnarray}
should be small compared with $V_1$ and $V_2$. This requires:
\beq
|{\bar \lambda}\! \cdot\!f |^2 \ \ll \ |\lambda|^2 f^2 \, .
\label{gf13}
\eeq
The condition~(\ref{gf13}) has a simple interpretation in terms
of goldstino couplings. The low-energy goldstino superfield
is defined by:
\beq
X \ \equiv \ \frac{1}{f} \ \sum_i f_i X_i \ ,
\label{gf14}
\eeq
such that, in the MSSM vacuum, $F_X = - f $ while the orthogonal
combinations $\chi_i$ ($i = 1 \cdots N-1$) have vanishing F-terms.
Denoting by $\lambda_X \equiv (\lambda\! \cdot\! \bar f) / f$ the coupling
of the messengers to the goldstino superfield, we can rewrite
the condition~(\ref{gf13}) in the simpler form:
\beq
|\lambda_X| \ \ll \ |\lambda| \ = \
\left( |\lambda_X|^2 + \sum_{i=1}^{N-1} |\lambda_{\chi_i}|^2 \right)^{1/2} .
\label{gf16}
\eeq
When Eq.~(\ref{gf16}), or equivalently Eq.~(\ref{gf14}), is satisfied,
the tree-level MSSM and messenger vacua are sufficiently
close in energy for quantum corrections to significantly affect
the vacuum structure of the theory.

\section{One-loop corrections to the vacuum energy }
\label{sec:oneloop}

We now turn to the explicit computation of the one-loop effective
potential in the subclass of models defined by the following
superpotential:
\begin{equation}
W \ = \ X_i \, ( f_i + \frac{1}{2} h_a^{(i)} \varphi_a^2) \, + \, m_a\, \varphi_a Y_a \,
+ \, \phi \, (\lambda\! \cdot\! X  + M ) \, \tilde{\phi} \ ,
\label{ol1}
\end{equation}
where $i = 1 \cdots Q$ and $a= 1 \cdots P$. The F-term equations
of motion read:
\begin{eqnarray}
&& - \bar F_{X_i}= f_i + \frac{1}{2} h_a^{(i)} \varphi_a^2 + \lambda_i \phi
\tilde{\phi} \ , \quad - \bar F_{Y_a} = m_a \varphi_a \nonumber \ , \\
&& - \bar F_{\varphi_a}= X_i h_a^{(i)} \varphi_a + m_a Y_a \ , \nonumber \\
&& - \bar F_{\phi} = (\lambda\! \cdot\! X  +M) \tilde{\phi} \ , \quad
- \bar F_{\tilde{\phi}} = (\lambda\! \cdot\! X  +M) \phi \ ,
\label{ol2}
\end{eqnarray}
and the tree-level scalar potential is given by:
\begin{eqnarray}
V & = & \sum_i\, \left| f_i + \frac{1}{2} h_a^{(i)} \varphi_a^2 + \lambda_i \phi
\tilde{\phi} \right|^2 +\, \sum_a\, | m_a \varphi_a|^2
+\, \sum_a \left| X_i h_a^{(i)} \varphi_a + m_a Y_a \right|^2 \nonumber \\
&& +\, \ | \lambda\! \cdot\! X  +M |^2 \ (|\tilde{\phi}|^2+ |\phi |^2) \ .
\label{ol3}
\end{eqnarray}
At tree level, $\langle \varphi_a \rangle = \langle Y_a \rangle = 0$
is realized for large enough values of $m_a$ (for instance, in the
MSSM vacuum the condition reads $m_a^2 > |\bar h_a\! \cdot\! f|$).
Supersymmetry is broken for any $Q > 1$. In the MSSM vacuum, the
$X_i$ fields are tree-level flat directions, whereas in the messenger
vacuum there are $Q-1$ flat directions if one imposes the
D-term constraint $\phi = {\tilde \phi}$.
The fermionic mass matrix has the general form:
\begin{eqnarray}
M_F\ =\
\begin{pmatrix}
{\cal M}_1 & 0 \\
0  & {\cal M}_2
\end{pmatrix} ,
\end{eqnarray}
where
\begin{eqnarray}
{\cal M}_{1}\ =\
\begin{pmatrix}
h_a^{(i)} X_i  & m_a \\
m_a   & 0
\end{pmatrix}
\end{eqnarray}
and
\begin{eqnarray}
{\cal M}_2\  =\
\begin{pmatrix}
0  & \lambda\! \cdot\!X + M  & \lambda_j {\tilde \phi} \\
\lambda\! \cdot\!X + M  & 0 & \lambda_j \phi  \\
\lambda_i {\tilde \phi}  & \lambda_i \phi & 0
\end{pmatrix}\, .
\label{ol4}
\end{eqnarray}
In order to compute the one-loop vacuum energies, we shall perform
the approximate calculation of the effective potential using the one-loop
K\"ahler potential~\cite{grisaru}:
\beq
K^{(1)} \ = \ - \ \frac{1}{32 \pi^2} \ {\rm Tr} \left( M_F M_F^{\dagger}\,
  \ln \frac{M_F M_F^{\dagger}}{\Lambda^2} \right) \, .
\label{ol5}
\eeq
Then the one-loop scalar potential is given by
\beq
V \ = \ (K^{-1})_{i j} \ F_{i} {\bar F}_{j} \ \equiv \ V_0 \ + \ V^{(1)} \ ,
\label{ol6}
\eeq
where at the linearized order in the corrections to the K\"ahler metric
we find:
\beq
V^{(1)} \ = \ \frac{1}{32 \pi^2} \ \sum_{\alpha}\,
\left[\, \frac{\partial^2 \mu_{\alpha}^2 }{{\partial X_i}{\partial {\bar X}_j} }
\left( \ln \frac{\mu_{\alpha}^2}{\Lambda^2}+1 \right)
+ \frac{1}{\mu_{\alpha}^2} \frac{\partial \mu_{\alpha}^2}{\partial X_i}
  \frac{\partial \mu_{\alpha}^2}{\partial {\bar X}_j}\, \right] F_{i} {\bar F}_{j} \ .
\label{ol6bis}
\eeq
In Eqs.~(\ref{ol6}) and~(\ref{ol6bis}), $(K^{-1})_{i j}$ is the inverse
of the K\"ahler metric
$K_{i j} = \frac{\partial^2 K}{\partial X_i \partial {\bar X}_j}$,
and $\mu_{\alpha}^2$ are the eigenvalues of $M_F M_F^{\dagger}$.

The eigenvalues of the mass matrix of the O'Raifeartaigh fields
$\varphi_a$ and $Y_a$, ${\cal M}_1$, are easily found. Writing:
\begin{eqnarray}
{\cal M}_1^a {\cal M}_1^{a\, \dagger}\ =\
\begin{pmatrix}
|h_a^{(i)} X_i |^2 + m_a^2 & m_a h_a^{(i)} X_i \\
m_a \bar h_a^{(i)} X_i^\dagger & m_a^2
\end{pmatrix} \, ,
\label{ol7}
\end{eqnarray}
one obtains the eigenvalues ($a = 1 \cdots P$):
\begin{eqnarray}
\mu_{a, \pm}^2 \ = \ \frac{1}{2} \left( 2 m_a^2 + |h_a^{(i)} X_i |^2
\pm |h_a^{(i)} X_i | \sqrt{|h_a^{(i)} X_i |^2 + 4 m_a^2} \right) \, ,
\label{ol8}
\end{eqnarray}
where, without loss of generality, the $m_a$ have been assumed
to be real parameters. The contribution of the $\varphi_a$, $Y_a$
fields to the effective K\"ahler potential is then:
\bea
&& \mbox{Tr}\, \left( {\cal M}_1 {\cal M}_1^\dagger \,
\ln\frac{{\cal M}_1 {\cal M}_1^\dagger}{\Lambda^2} \right)\, =\
\sum_a\ \Biggl\{ \left(|h_a^{(i)} X_i |^2 + 2 m_a^2 \right)
\ln\frac{m_a^2}{\Lambda^2} \nonumber \\
&& \hskip 1cm +\ 2\ |h_a^{(i)} X_i |\, \sqrt{|h_a^{(i)} X_i |^2 + 4 m_a^2} \
\ln\frac{|h_a^{(i)} X_i | + \sqrt{|h_a^{(i)} X_i |^2 + 4 m_a^2} }{2 m_a} \ \Biggr\} \ .
\label{ol9}
\eea

In the absence of messenger fields, the one-loop effective potential
can be easily analyzed in the K\"ahler approximation (in the
small supersymmetry breaking limit). In this case the fermion mass
matrix reduces to ${\cal M}_1$, and the K\"ahler metric is given by:
\beq
K_{i j} \ = \ \delta_{ij} + Z_a h_a^{(i)} {\bar h}_a^{(j)}\ ,
\label{ol10}
\eeq
where
\bea
&& Z_a \ = \ - \frac{1}{32 \pi^2}\ \Biggl\{\, \ln\frac{m_a^2}{\Lambda^2}\,
+\, 2\, -\, \frac{2 m_a^2 }{|h_a^{(i)} X_i |^2 + 4 m_a^2}  \nonumber  \\
&& +\ \frac{2}{|h_a^{(i)} X_i |}\,
\frac{|h_a^{(i)} X_i |^4 + 6 m^2_a |h_a^{(i)} X_i |^2 + 4 m_a^4}
{(|h_a^{(i)} X_i |^2 + 4 m^2)^{3/2}}\
\ln \frac{|h_a^{(i)} X_i |+\sqrt{ |h_a^{(i)}X_i |^2 + 4 m_a^2}}{2 m_a}\ \Biggl\}  \ .
\hskip 1cm
\label{ol11}
\eea
Let us define $\chi_a \equiv h_a^{(i)} X_i$. The functions $Z_a$
are monotonically decreasing functions of $|\chi_a|$, whose limiting
values are given by:
\bea
&& Z_a\, (|\chi_a| \ll m_a) \ \simeq \ - \ \frac{1}{32 \pi^2}
\left( 2 + \ln \frac{m_a^2}{\Lambda^2} + \frac{2 |\chi_a|^2}{3 m_a^2}  \right) ,
\nonumber \\
&& Z_a\, (|\chi_a| \gg m_a)  \ \simeq \
- \ \frac{1}{32 \pi^2}\, \ln \frac{|\chi_a|^2}{\Lambda^2} \ .
\label{ol12}
\eea
Since $Z_a \ll 1$, the inverse K\"ahler metric is simply:
\beq
K^{-1}_{i j} \ = \ \delta_{ij} \ - \ Z_a \ h_a^{(i)} {\bar h}_a^{(j)} \, ,
\label{ol13}
\eeq
and the one-loop effective potential is:
\beq
V^{(1)} \ = \ - \ Z_a (|\chi_a|) \ |\bar h_a\! \cdot\! f |^2\, .
\label{ol14}
\eeq
The effect of the one-loop corrections is to lift the tree-level flat
directions and to stabilize the pseudo-moduli fields $X_i$ at the origin.
More precisely, all $X_i$'s are stabilized at the origin for $P \geq Q$
if the couplings $h^{(i)}_a$ are generic, while some flat directions
are still present for $P < Q$. This can easily be seen by expanding
the effective potential~(\ref{ol14}) for small $X_i$ values:
\beq
V^{(1)}\ \simeq\ {\rm const}\, +\, \ \frac{1}{32 \pi^2}
\left( 2 + \ln \frac{m_a^2}{\Lambda^2}
+ \frac{2}{3 m_a^2}\, \bar h_a^{(i)} h_a^{(j)} \bar X_i X_j \right)
|\bar h_a\! \cdot\! f |^2 \, .
\label{ol15}
\eeq
All pseudo-moduli fields are stabilized at $X_i = 0$
if the positive matrix
\beq
M_{ij}^2 \ \equiv \ \sum_a\, \frac{|\bar h_a\! \cdot\! f |^2 }{m_a^2}\,
\bar h_a^{(i)} h_a^{(j)} \,
\label{ol16}
\eeq
has rank $Q$. For generic $h^{(i)}_a$ couplings, this is the case
for $P \geq Q$ (notice that in the case $P>Q$ the fields $\chi_a$
are not independent of each other). One should also keep in mind
that the constraint $m_a^2 > |h_a\! \cdot\!{\bar f}|$ has to be
imposed in order to stabilize the fields $\varphi_a$ and $Y_a$
at the origin.

Let us now reintroduce the messenger fields. The second part
of the fermion mass matrix, coming from the messenger fields, gives:
\begin{eqnarray}
{\cal M}_2 {\cal M}_2^{\dagger}\, =
\begin{pmatrix}
| \lambda\! \cdot\! X  +M |^2 + |\lambda|^2 |\tilde{\phi}|^2
  & |\lambda|^2 \tilde{\phi} \phi^\dagger
  & \bar \lambda_j \phi^\dagger (\lambda\! \cdot\!X +M)  \\
|\lambda|^2 \phi \tilde{\phi}^\dagger
  & | \lambda\! \cdot\! X  +M |^2  + |\lambda|^2 |\phi|^2
  & \bar \lambda_j \tilde{\phi}^\dagger (\lambda\! \cdot\!X +M) \\
\lambda_i \phi\, (\lambda\! \cdot\!X +M)^\dagger
  & \lambda_i \tilde\phi\, (\lambda\! \cdot\!X +M)^\dagger
  & \lambda_i \bar \lambda_j (|\phi|^2 + |\tilde{\phi}|^2)
\end{pmatrix} .
\label{ol17}
\end{eqnarray}
It can be shown that this matrix has $Q-1$ zero eigenvalues,
corresponding to $Q-1$ tree-level flat directions present both in the
MSSM and in the messenger vacuum. The remaining eigenvalues
are the solutions of the following equation:
\beq
\mu^2\, \left(\mu^2 - |\lambda|^2 (|\phi|^2 + |\tilde{\phi}|^2)
- |\lambda\! \cdot\!X +M |^2 \right)^2 =\
4\, |\lambda|^4\, |\phi \tilde{\phi}|^2\, |\lambda\! \cdot\!X +M |^2\, .
\label{ol18}
\eeq
In the MSSM vacuum, $\phi = \tilde \phi = 0$ and another zero
eigenvalue is found, corresponding to the Q$^{\rm th}$ flat direction
of the tree-level scalar potential.

\subsection{One-loop corrections to the MSSM vacuum energy}

Due to the vanishing messenger vevs in the MSSM vacuum,
the matrix ${\cal M}_2 {\cal M}_2^\dagger$ has only two equal
nonzero eigenvalues $\mu^2 = |\lambda\! \cdot\!X + M|^2$. Hence:
\beq
\mbox{Tr} \left( {\cal M}_2 {\cal M}_2^\dagger\,
\ln \frac{{\cal M}_2 {\cal M}_2^\dagger}{\Lambda^2} \right)\,
=\ 2\, |\lambda\! \cdot\! X + M|^2\,
\ln \frac{|\lambda\! \cdot\!X + M|^2}{\Lambda^2} \ .
\label{sm1}
\eeq
Putting all contributions together, the K\"ahler metric is given by:
\beq
K_{i j} \ = \ \delta_{ij} + Z_a h_a^{(i)} {\bar h}_a^{(j)}
+ Z' \lambda_i {\bar \lambda}_j \ ,
\label{sm2}
\eeq
where the functions $Z_a (|\chi_a|)$ are given by Eq.~(\ref{ol11})
as before, and
\beq
Z'\ =\ -\frac{1}{16 \pi^2}\,
\Bigg( \mbox{ln}|\lambda\! \cdot\!X +M|^2 +2 \Bigg) \ .
\label{sm3}
\eeq
In order to be able to write some analytic minimization conditions,
let us assume that the pseudo-moduli fields $X_i$ are stabilized
close to the origin, namely that $|X_i| \ll m_a, M$ (later on we will
derive condition for this to be the case).
We can then expand the one-loop effective potential
\beq
V^{(1)} \ \simeq \ -\, Z_a\, |{\bar h}_a\! \cdot\! f|^2 \,
- \, Z'\, |\bar{\lambda}\! \cdot\!f |^2 \, ,
\label{sm4}
\eeq
and, for $P \geq Q$, we find a minimum at:
\beq
M_{ij}^2 X_j \ = \ - \, \frac{3 \bar \lambda_i}{M} \
|\bar{\lambda}\! \cdot\!f |^2 \, ,
\label{sm5}
\eeq
where the matrix $M_{ij}^2$ has been defined in Eq.~(\ref{ol16}).
The pseudo-moduli fields $X_i$ are therefore stabilized at small
values $|X_i| \ll m_a, M$ as soon as $m_a \ll M$ (or even
$m_a < M$ if the couplings $h^{(i)}_a$ are of order $1$),
implying that the messengers cannot be too light.
Setting $X_i = 0$ in the effective potential~(\ref{sm4}) then gives
a very good approximation of the one-loop MSSM vacuum energy:
\beq
V_1 \ = \ f^2\, +\, \frac{1}{32 \pi^2} \left[\, \sum_a |\bar h_a\! \cdot\! f|^2
\left( \ln \frac{m_a^2}{\Lambda^2} + 2 \right)
+\, 2\, |\bar{\lambda}\! \cdot\!f |^2 \left( \ln \frac{M^2}{\Lambda^2} + 2 \right) \right] \, .
\label{sm6}
\eeq
Using the renormalization group equations of Appendix~\ref{app:RGEs},
it is easy to show that the $\ln \Lambda$-dependent terms in $V_1$
are precisely renormalizing the tree-level vacuum energy $f^2$.
One can therefore write:
\beq
V_1 \ = \ f^2 (\mu)\, +\, \frac{1}{32 \pi^2} \left[\, \sum_a |\bar h_a\! \cdot\! f|^2
\left( \ln \frac{m_a^2}{\mu^2} + 2 \right)
+\, 2\, |\bar{\lambda}\! \cdot\!f |^2 \left( \ln \frac{M^2}{\mu^2} + 2 \right) \right] \, ,
\label{sm7}
\eeq
where the couplings in Eq.~(\ref{sm7}) are evaluated at the renormalization
group scale $\mu$.

\subsection{One-loop corrections to the messenger vacuum energy}
\label{subsec:olmv}

In the messenger vacuum, one has:
\beq
\lambda\! \cdot\! X  +M \, = \, 0\  , \qquad \phi \tilde{\phi} \
= \, - \frac{1}{|\lambda|^2}\, \bar \lambda\! \cdot\! f \ (1 + \epsilon_{\phi} )\ ,
\label{m1}
\eeq
where $\epsilon_{\phi}$ is a one-loop correction
to the tree-level messenger vevs.
In Eq.~(\ref{m1}) we anticipated the fact that $\lambda\! \cdot\!X + M = 0$
is also valid at the one-loop level, since there are no anomalous
dimensions mixing the messenger fields with the O'Raifeartaigh fields
($\gamma_{X_i}^{\phi} = \gamma_{X_i}^{\tilde \phi} =0$). Thus,
$F_{\phi} = F_{\tilde \phi} = 0$ still holds at the one-loop level, and since
$\bar F_{\phi} = - (\lambda\! \cdot\!X + M)\, {\tilde \phi}$ this also implies
$\lambda\! \cdot\!X + M = 0$.
However, in order to keep the full $X_i$-dependence of the
one-loop effective potential, one must solve Eq.~(\ref{ol18}) for
$\lambda\! \cdot\!X + M \neq 0$. This can be done in the limit
$|\lambda\! \cdot\!X +M|^2 \ll |\lambda|^2 (|\phi|^2 + |\tilde{\phi}|^2)$,
in which:
\begin{eqnarray}
\mu_1^2 \!\! & \simeq\! &\! \frac{4 |\phi \tilde{\phi}|^2 }{(|\phi| ^2+ |\tilde{\phi}|^2)^2}\,
|\lambda\! \cdot\!X + M |^2\, , \label{m2}  \\
\mu_{2,3}^2 \!\! & \simeq\! &\! |\lambda|^2 (|\phi| ^2+ |\tilde{\phi}|^2)
\pm \frac{2\, |\lambda \phi \tilde{\phi}|}{(|\phi|^2 + |\tilde{\phi}|^2)^{1/2}}\,
|\lambda\! \cdot\!X+ M| + \frac{|\phi|^4+ |\tilde{\phi}|^4}{(|\phi|^2+ |\tilde{\phi}|^2)^2}\,
|\lambda\! \cdot\!X+M|^2 \, . \hskip .5cm \nonumber
\end{eqnarray}
The K\"ahler metric then reads:
\beq
K_{i j} \ = \ \delta_{ij} + Z_a h_a^{(i)} {\bar h}_a^{(j)}
+ Z' \lambda_i {\bar \lambda}_j \ ,
\label{m3}
\eeq
where
\bea
Z'\!  & =\! & -\, \frac{1}{16 \pi^2}\, -\, \frac{1}{16 \pi^2 (|\phi|^2 + |\tilde{\phi}|^2)^2}\
\Bigg\{ 2\, |\phi \tilde{\phi}|^2 \left(
\ln\, \frac{4\, |\phi \tilde{\phi}|^2\, |\lambda\! \cdot\!X +M |^2}
{(|\phi|^2 + |\tilde{\phi}|^2)^2\, \Lambda^2}\, +2\right) \nonumber \\
&& +\ (|\phi|^4 + |\tilde{\phi}|^4)\, \ln\, \frac{|\lambda|^2\, (|\phi|^2 + |\tilde{\phi}|^2)}
{\Lambda^2}\, \Bigg\} \ .
\label{m4}
\eea
Since $\lambda_i F_i=0$ in the tree-level messenger vacuum,
the term proportional to $Z'$ contributes to the effective potential
only at higher loop level. Hence, the one-loop effective potential
reduces to:
\beq
V^{(1)} \ \simeq \ - \sum_a Z_a \,
|\bar h_a\! \cdot\! f + (\bar h_a\! \cdot\! \lambda)\, \phi \tilde{\phi}|^2 \, .
\label{m5}
\eeq
Minimization of Eq.~(\ref{m5}) with respect to $\phi$, $\tilde \phi$
confirms that the messenger vevs are of the expected form~(\ref{m1})
and yields the one-loop vacuum energy\footnote{Since $V_2$
corresponds to a stationary point of the scalar potential, it does
not depend linearly on the one-loop correction $\epsilon_\phi$.
Terms quadratic in $\epsilon_\phi$ would be formally two-loop
and have been omitted.}:
\beq
V_2\, (\chi_a)\ =\ \sum_i\, \left| f_i - \frac{\lambda_i}{|\lambda|^2}\,
{\bar \lambda}\! \cdot\!f \right|^2 -\, \sum_a Z_a
\left| {\bar h}_a\! \cdot\!f - ({\bar h}_a\! \cdot\!\lambda)\,
\frac{{\bar \lambda}\! \cdot\!f}{|\lambda|^2} \right|^2 \, ,
\label{m6}
\eeq
where the minimization with respect to the $X_i$ fields remains
to be done.
Each function $Z_a (|\chi_a|)$ is separately minimized for $\chi_a = 0$.
However, for $P \geq Q$ the $\chi_a$'s are not independent variables,
so that it is not possible to set all of them to zero. Hence the one-loop
messenger vacuum energy will in general be larger than
$V_2 (\chi_a=0)$:
\beq
V_2 \ > \ \sum_i\, \left| f_i - \frac{\lambda_i}{|\lambda|^2}\,
{\bar \lambda}\! \cdot\!f \right|^2 +\, \frac{1}{32 \pi^2}\,
\sum_a \left( \ln \frac{m_a^2}{\Lambda^2} + 2 \right)
\left| {\bar h}_a\! \cdot\!f - ({\bar h}_a\! \cdot\!\lambda)\,
\frac{{\bar \lambda}\! \cdot\!f}{|\lambda|^2} \right|^2 \, .
\label{m7}
\eeq
Using the renormalization group equations of Appendix~\ref{app:RGEs},
one can show that the $\ln \Lambda$-dependent term in $V_2$, which
has exactly the same form as the one in the RHS of Eq.~(\ref{m7}),
renormalizes the tree-level vacuum energy. One can therefore write:
\beq
V_2 \ > \ \sum_i\, \left| f_i - \frac{\lambda_i}{|\lambda|^2}\,
{\bar \lambda}\! \cdot\!f \right|^2\! (\mu)\, +\, \frac{1}{32 \pi^2}\,
\sum_a \left( \ln \frac{m_a^2}{\mu^2} + 2 \right)
\left| {\bar h}_a\! \cdot\!f - ({\bar h}_a\! \cdot\!\lambda)\,
\frac{{\bar \lambda}\! \cdot\!f}{|\lambda|^2} \right|^2 \, .
\label{m8}
\eeq
where the couplings in Eq.~(\ref{m8}) are evaluated at the
renormalization group scale $\mu$.

\subsection{Comparison of the one-loop energies of the two vacua}

Let us now write the condition for the one-loop energy of the MSSM
vacuum to be lower than the one of the messenger vacuum.
Using Eqs.~(\ref{sm7}) and~(\ref{m8}), we obtain the following
upper bound on $\Delta V \equiv V_1 - V_2$:
\bea
\Delta V\! & <\! & |{\bar \lambda}\! \cdot\!f|^2\, \Biggl\{ \frac{1}{|\lambda|^2}
+ \frac{1}{32 \pi^2} \left[\, 2 \left( \ln \frac{M^2}{\mu^2} + 2 \right)
- \frac{1}{|\lambda|^4} \sum_a \left( \ln \frac{m_a^2}{\mu^2} + 2 \right)
|{\bar h}_a\! \cdot\!\lambda|^2\, \right]  \Biggr\} \nonumber  \\
&& + \ \frac{1}{32 \pi^2 |\lambda|^2} \, \sum_a
\left( \ln \frac{m_a^2}{\mu^2} + 2 \right)
\left[\, (h_a\! \cdot\!{\bar f}) ({\bar h}_a\! \cdot\!\lambda) ({\bar \lambda}\! \cdot\!f)
+ \mbox{c.c.}\, \right] \, .
\label{c1}
\eea
The first line in Eq.~(\ref{c1}) is dominated by the tree-level term and
is therefore positive, while the second line does not have a definite
sign. Since the latter is proportional to ${\bar \lambda}\! \cdot\!f$,
it can overcome the former, which is proportional to
$|{\bar \lambda}\! \cdot\!f|^2$ (remember that we have required
$|{\bar \lambda}\! \cdot\!f|^2 \ll |\lambda|^2 f^2$), and promote
the MSSM vacuum to the ground state of the theory. For this
to happen, a sufficient condition is that the superpotential parameters
be such that the RHS of Eq.~(\ref{c1}) is negative. This condition
simplifies in the case of equal O'Raifeartaigh masses $m_a = m$ to:
\bea
|{\bar \lambda}\! \cdot\!f|^2\ \Biggl\{ 1
+ \frac{1}{16 \pi^2} \left[\, |\lambda|^2 \left( \ln \frac{M^2}{m^2} + 2 \right)
- \sum_a\, \frac{|{\bar h}_a\! \cdot\!\lambda|^2}{|\lambda|^2}\, \right]  \Biggr\}
\hskip 3cm \nonumber  \\
< \ -\, \frac{1}{8 \pi^2}\ \mbox{Re} \left[\,  ({\bar \lambda}\! \cdot\!f)
\sum_a\, (h_a\! \cdot\!{\bar f}) ({\bar h}_a\! \cdot\!\lambda)\, \right] \, ,
\label{c2}
\eea
with all couplings evaluated at the renormalization group
scale $\mu = m$. Neglecting the terms suppressed by a loop
factor in the LHS of Eq.~(\ref{c2}), one arrives at the simpler,
approximate condition (to be supplemented with
the appropriate condition on the coupling phases):
\beq
|{\bar \lambda}\! \cdot\!f| \  <  \ \frac{1}{8 \pi^2}\,
\left|\, \sum_a  (\lambda\! \cdot\!{\bar h}_a) (h_a\! \cdot\!{\bar f})\, \right| \ ,
\label{c3}
\eeq
which is the main result of this letter. In terms of the couplings
of the low-energy goldstino superfield, the same condition
is expressed by Eq.~(\ref{i2}).

Let us now summarize all the requirements we imposed on the
superpotential parameters in order to arrive at Eq.~(\ref{c3}):
\beq
|{\bar \lambda}\! \cdot\!f|\, \ll\, |\lambda| f \ ,
\qquad m_a\, \ll\, M \ , \qquad |{\bar h}_a\! \cdot\!f|\, <\, m_a^2 \ ,
\qquad |{\bar \lambda}\! \cdot\!f|\, <\, M^2\ .
\label{c4}
\eeq
The first inequality is the condition imposed on the difference
of the tree-level MSSM and messenger vacuum energies, and
is no longer needed once Eq.~(\ref{c3}) is satisfied\footnote{For
$h^{(i)}_a$ couplings at most of order one, as required by
perturbativity, Eq.~(\ref{c3}) actually implies
$|{\bar \lambda}\! \cdot\!f| \ll |\lambda| f$, or a weaker form of it.}.
The second inequality ensures that the pseudo-moduli fields
$X_i$ are stabilized close to the origin in the MSSM vacuum,
a fact that was taken into account in the computation of the
MSSM vacuum energy. The last two inequalities are required
to avoid the presence of tachyons in the O'Raifeartaigh and
messenger sectors, respectively (the fourth one is actually
an automatic consequence of the other constraints, which
even imply $|{\bar \lambda}\! \cdot\!f| \ll M^2$).

\section{Final comments}

Let us review the assumptions made in the derivation of the
condition~(\ref{c3}). First it was obtained in a specific
class of perturbative supersymmetry breaking models coupled
to messenger fields. The computation of the vacuum energies
was limited to the one-loop level and made in the K\"ahler
approximation (this is however legitimate in the limit of small
supersymmetry breaking, $|{\bar h}_a\! \cdot\!f| \ll m_a^2$
and $|{\bar \lambda}\! \cdot\!f| \ll M^2$). Finally, the validity of
our one-loop computation is strictly speaking limited to the vicinity
of the tree-level vacua, and we cannot exclude the presence
of other minima in the one-loop scalar potential, although we
view this as a rather unlikely possibility.
All in all we believe that, while they do not constitute a rigourous
proof, our computations and arguments provide strong evidence
that quantum corrections can make the MSSM vacuum absolutely
stable, even though instabilities in the direction of the messenger
fields are present at tree level. An important point is that gaugino
masses are not suppressed relative to soft scalar masses, in
contrast to the tree-level supersymmetry breaking models
discussed in Ref.~\cite{ks}.

Finally, we would like to comment on the constraints set by
Eqs.~(\ref{c3}) and~(\ref{c4}) on the mass scales involved in the class
of models we have considered. Imposing a perturbative upper
bound of order $1$ on dimensionless parameters, we obtain
$|{\bar \lambda}\! \cdot\!f| \lesssim m^2_a / (8 \pi^2)$ and
$|{\bar \lambda}\! \cdot\! f| \ll M^2 / (8 \pi^2)$. Since MSSM
soft terms in the few $100 \GeV$ -- $1 \TeV$ range require
$|{\bar \lambda}\! \cdot\! f| / M \sim 100 \TeV$ in perturbative
gauge mediation, heavy O'Raifeartaigh and~messenger fields
are required:
\beq
M\ \gg\ m_a\ \gtrsim\ \sqrt{(10^4\, \TeV) M}\ .
\label{fc1}
\eeq
The minimal allowed values for the various mass scales involved
are:
\beq
m_a\, \sim\, 10^5 \TeV\ , \quad M\, \sim\, 10^6 \TeV\ , \quad
f\, \sim\, \lambda_X^{-1} \, (10^4 \TeV)^2 \ ,
\label{fc2}
\eeq
where $\lambda_X = (\lambda\! \cdot\! \bar f) / f$ is the 
goldstino-messenger coupling.  As for
the masses of the pseudo-moduli fields $X_i$, they are given by
the eigenvalues of the matrix $M^2_{ij}$ defined in Eq.~(\ref{ol16})
and do not possess a model-independent lower bound;
for $h^{(i)}_a \sim 1$ and $\lambda_X \sim 10^{-2}$, they
are of order $10^5 \TeV$. All these states are well beyond the
reach of high-energy colliders.
Notice that the lowest achievable gravitino mass is of order
$m_{3/2} \sim 10^{-2} \GeV$ (corresponding to $\lambda_X \sim 10^{-2}$), 
which allows to evade the most severe BBN problems associated
with NLSP decays. Such a gravitino mass is also consistent
with gravitino as cold dark matter. 

We conclude that heavy messenger and supersymmetry breaking
fields seem to be required in order for one-loop corrections to ensure
the stability of the MSSM vacuum.


\section*{Acknowledgments}
{This work was partially supported by the European ERC Advanced
Grant 226371 MassTeV, by the CNRS PICS Nos. 3747 and 4172,
by the European Initial Training Network PITN-GA-2009-237920
and by the Agence Nationale de la Recherche.}


\begin{appendix}

\renewcommand{\theequation}{A.\arabic{equation}}
\setcounter{equation}{0}  

\section{Renormalization group equations}
\label{app:RGEs}

The renormalization group equations for the superpotential
couplings~(\ref{ol1}) are:
\bea
\frac{d}{dt} \ h_a^{(i)}\! & =\! & \frac{1}{16 \pi^2}\, \left[\, 2  |h|^2 h_a^{(i)}
+ \frac{1}{2}\, h_b^{(i)} {\bar h}_b^{(j)} h_a^{(j)}
+ (h_a\! \cdot\! {\bar \lambda}) \lambda_i\,  \right] \, ,
\label{a1}  \\
\frac{d}{dt} \ \lambda_i\! & =\! & \frac{1}{16 \pi^2}\, \left[\, 3 |\lambda|^2 \lambda_i
+ \frac{1}{2}\, ({\bar h}_b\! \cdot\! \lambda)  h_b^{(i)}\, \right] \, ,
\label{a2}  \\
\frac{d}{dt} \ f_i\! & =\! & \frac{1}{16 \pi^2}\, \left[\, \frac{1}{2}\, h_a^{(i)} {\bar h}_a^{(j)}
+ \lambda_i {\bar \lambda}_j\, \right] f_j \ ,
\label{a3}
\eea
where $|h|^2 \equiv \sum_a \sum_i h_a^{(i)} {\bar h}_a^{(i)}$.


\renewcommand{\theequation}{B.\arabic{equation}}
\setcounter{equation}{0}  

\section{More about the messenger vacuum}
\label{app:inversion}

In Section~\ref{subsec:olmv}, we argued that the term proportional
to $Z'$ contributes to the effective potential only at higher loop level,
since $\lambda_i F_i=0$ in the tree-level messenger vacuum.
However, since the $Z'$ function~(\ref{m4}) diverges for
$\lambda\! \cdot\!X + M \to 0$ one can wonder whether
it is legitimate to do so. In this appendix, we propose an alternative
computation of the one-loop effective potential in the messenger
vacuum, based on the exact inversion of the K\"ahler metric,
which supports the result of Section~\ref{subsec:olmv}.

Starting from the K\"ahler metric:
\beq
K_{ij} \ = \ \delta_{ij} + Z_a h_a^{(i)} \bar{h}_a^{(j)} + Z' \lambda_i \bar{\lambda}_j \ ,
\label{b1}
\eeq
we can formally invert it exactly into:
\bea
(K^{-1})_{jk} & = &\delta_{jk} - Z_b\, h_a^{(j)} N^{-1}_{ab} \bar{h}^k_b
- \frac{Z' Z_c Z_d}{1+|\lambda|^2 Z'}\, (\lambda\! \cdot\!{\bar h}_c)
(\bar{\lambda}\! \cdot\!h_d)\, h_a^{(j)} N^{-1}_{ab}
(M^{-1})^\dagger_{db} {\bar h}_b^{(k)}  \nonumber  \\
&& + \ \frac{Z' Z_b}{1+|\lambda|^2 Z'}
\left[\, h_a^{(j)}M^{-1}_{ab} (\bar{h}_b\! \cdot\!\lambda) \bar{\lambda}_k
+ \lambda_j (\bar{\lambda}\! \cdot\!h_b) (M^{-1})^\dagger_{ba} h_a^{(k)}\, \right]
\nonumber \\
&&  - \ \frac{Z'}{1+|\lambda|^2 Z'} \left[\,
1 + \frac{Z'Z_b}{1+|\lambda|^2 Z'} \, (\bar{\lambda}\! \cdot\!h_a)
({\bar h}_b\! \cdot\!\lambda ) M^{-1}_{ab}\, \right] \lambda_j \bar{\lambda}_k \ ,
\label{b2}
\eea
where the matrices $M$ and $N$ are defined by:
\bea
M_{ab}\! & =\! & \delta_{ab}\, +\, Z_a\, {\bar h}_a\! \cdot\!h_b\,
-\, \frac{Z_a Z'}{1 + |\lambda|^2 Z'} \, ({\bar h}_a\! \cdot\!\lambda)
({\bar \lambda}\! \cdot\!h_b) \ , \nonumber \\
N_{ab}\! & =\! & \delta_{ab}\, +\, Z_a\,  {\bar h}_a\! \cdot\!h_b \ .
\label{b3}
\eea
In the limit $Z' \gg 1$, $Z_a \ll 1$, we obtain:
\bea
(K^{-1})_{jk} \ = \ \delta_{jk}\, -\, \frac{\lambda_j \bar{\lambda}_k}{|\lambda|^2}\,
-\, Z_a\, \Bigg[\, h_a^{(j)} \bar{h}_a^{(k)}\,
-\, \frac{\bar{h}_a\! \cdot\! \lambda}{|\lambda|^2}\ h_a^{(j)} \bar \lambda_k\,
-\, \frac{h_a\! \cdot\! \bar{\lambda}}{|\lambda|^2}\ \bar{h}_a^{(k)}  \lambda_j
\hskip 1cm  \nonumber  \\
+\ \frac{(\bar{\lambda}\! \cdot\!h_a)(\bar{h}_a\! \cdot\!\lambda)}{|\lambda|^4}\
\lambda_j \bar{\lambda}_k\,  \Bigg]\, +\ {\cal O} \left( \frac{1}{Z'}\, , Z_a \right)\, .
\label{b4}
\eea
Thus, even though $Z'$ diverges, the inverse K\"ahler metric remains
finite. It is interesting to note that taking the limit $Z' \to \infty$ leaves
the term $- \lambda_j \bar{\lambda}_k / |\lambda|^2$
in Eq.~(\ref{b4}), which is not suppressed by a loop factor.
Now the one-loop vacuum energy reads:
\bea
V_2\, (\chi_a) & = & (K^{-1})_{i j} \ F_{i} {\bar F}_{j}  \nonumber  \\
& = & f^2 \, - \, \frac{1}{|\lambda|^2}\, |\bar{\lambda}\! \cdot\!f |^2\,
-\, Z_a \left|\, \bar{h}_a\! \cdot\!f - (\bar{h}_a\! \cdot\!\lambda)\,
\frac{\bar{\lambda}\! \cdot\!f }{|\lambda|^2}\, \right|^2 \, ,
\label{b5}
\eea
(where we have inserted the tree-level messenger vevs
$\phi \tilde{\phi} = - \bar \lambda\! \cdot\! f / |\lambda|^2$),
in agreement with Eq.~(\ref{m6}).

\end{appendix}



\begin{thebibliography}{99}

\bibitem{gr} G.~F.~Giudice and R.~Rattazzi,
  Phys.\ Rept.\  {\bf 322} (1999) 419
  [arXiv:hep-ph/9801271].

\bibitem{cfs}
  C.~Cheung, A.~L.~Fitzpatrick and D.~Shih,
  JHEP {\bf 0807} (2008) 054
  [arXiv:0710.3585 [hep-ph]];
  P.~Meade, N.~Seiberg and D.~Shih,
  Prog.\ Theor.\ Phys.\ Suppl.\  {\bf 177}, 143 (2009)
  [arXiv:0801.3278 [hep-ph]];
  L.~M.~Carpenter, M.~Dine, G.~Festuccia and J.~D.~Mason,
  Phys.\ Rev.\  D {\bf 79}, 035002 (2009)
  [arXiv:0805.2944 [hep-ph]];
  H.~Ooguri, Y.~Ookouchi, C.~S.~Park and J.~Song,
  Nucl.\ Phys.\  B {\bf 808}, 121 (2009)
  [arXiv:0806.4733 [hep-th]];
  J.~Distler and D.~Robbins,
  arXiv:0807.2006 [hep-ph];
  K.~A.~Intriligator and M.~Sudano,
  JHEP {\bf 0811}, 008 (2008)
  [arXiv:0807.3942 [hep-ph]];
  K.~Benakli and M.~D.~Goodsell,
  Nucl.\ Phys.\  B {\bf 816}, 185 (2009)
  [arXiv:0811.4409 [hep-ph]];
  M.~Buican, P.~Meade, N.~Seiberg and D.~Shih,
  JHEP {\bf 0903}, 016 (2009)
  [arXiv:0812.3668 [hep-ph]];
  M.~Buican and Z.~Komargodski,
  JHEP {\bf 1002}, 005 (2010)
  [arXiv:0909.4824 [hep-ph]].

\bibitem{dlp}
  E.~Dudas, S.~Lavignac and J.~Parmentier,
  Nucl.\ Phys.\  B {\bf 808} (2009) 237
  [arXiv:0808.0562 [hep-ph]];
N.~J.~Craig and D.~R.~Green,
  Phys.\ Rev.\  D {\bf 79} (2009) 065030
  [arXiv:0808.1097 [hep-ph]];
M.~Ibe and T.~T.~Yanagida,
  Phys.\ Rev.\  D {\bf 81} (2010) 035017
  [arXiv:0912.4221 [hep-ph]].

\bibitem{dkk}
D.~Green, A.~Katz and Z.~Komargodski,
  arXiv:1008.2215 [hep-th];
M.~McGarrie and R.~Russo,
  Phys.\ Rev.\  D {\bf 82} (2010) 035001
  [arXiv:1004.3305 [hep-ph]];
M.~McGarrie,
  arXiv:1009.0012 [hep-ph];
 M.~Sudano,
  arXiv:1009.2086 [hep-ph].

\bibitem{abel}
S.~Abel, M.~J.~Dolan, J.~Jaeckel and V.~V.~Khoze,
  JHEP {\bf 0912} (2009) 001
  [arXiv:0910.2674 [hep-ph]];
S.~Shirai, M.~Yamazaki and K.~Yonekura,
  JHEP {\bf 1006} (2010) 056
  [arXiv:1003.3155 [hep-ph]];
S.~Abel, M.~J.~Dolan, J.~Jaeckel, V.~V.~Khoze, M.~J.~Dolan, J.~Jaeckel and V.~V.~Khoze,
  arXiv:1009.1164 [hep-ph].

\bibitem{gaugepheno}
S.~Dimopoulos, S.~D.~Thomas and J.~D.~Wells,
  Nucl.\ Phys.\  B {\bf 488} (1997) 39
  [arXiv:hep-ph/9609434];
J.~A.~Bagger, K.~T.~Matchev, D.~M.~Pierce and R.~j.~Zhang,
  Phys.\ Rev.\  D {\bf 55} (1997) 3188
  [arXiv:hep-ph/9609444].

\bibitem{Drees_book}
M.~Drees, R.~Godbole and P.~Roy,
  ``Theory and phenomenology of sparticles: An account of four-dimensional N=1
  supersymmetry in high energy physics'',
{\it  Hackensack, USA: World Scientific (2004) 555 p};
P.~Binetruy,
  ``Supersymmetry: Theory, experiment and cosmology,''
{\it  Oxford, UK: Oxford Univ. Pr. (2006) 520 p}. 

\bibitem{is}
K.~A.~Intriligator and N.~Seiberg,
  Class.\ Quant.\ Grav.\  {\bf 24} (2007) S741
  [arXiv:hep-ph/0702069].

\bibitem{gauge2}
M.~Dine, A.~E.~Nelson and Y.~Shirman,
  Phys.\ Rev.\  D {\bf 51} (1995) 1362
  [arXiv:hep-ph/9408384];
M.~Dine, A.~E.~Nelson, Y.~Nir and Y.~Shirman,
  Phys.\ Rev.\  D {\bf 53} (1996) 2658
  [arXiv:hep-ph/9507378].

\bibitem{ISS06}
K.~Intriligator, N.~Seiberg and D.~Shih,
  JHEP {\bf 0604} (2006) 021
  [arXiv:hep-th/0602239].

\bibitem{direct}
R.~Kitano, H.~Ooguri and Y.~Ookouchi,
  Phys.\ Rev.\  D {\bf 75} (2007) 045022
  [arXiv:hep-ph/0612139];
C.~Csaki, Y.~Shirman and J.~Terning,
  JHEP {\bf 0705} (2007) 099
  [arXiv:hep-ph/0612241];
S.~Abel, C.~Durnford, J.~Jaeckel and V.~V.~Khoze,
  Phys.\ Lett.\  B {\bf 661} (2008) 201
  [arXiv:0707.2958 [hep-ph]].
N.~Haba and N.~Maru,
  Phys.\ Rev.\  D {\bf 76} (2007) 115019
  [arXiv:0709.2945 [hep-ph]];
A.~Giveon, A.~Katz and Z.~Komargodski,
  JHEP {\bf 0907} (2009) 099
  [arXiv:0905.3387 [hep-th]];
  Y.~Nakai and Y.~Ookouchi,
  JHEP {\bf 1101} (2011) 093
  [arXiv:1010.5540 [hep-th]].

\bibitem{ray}
 S.~Ray,
  Phys.\ Lett.\  B {\bf 642} (2006) 137
  [arXiv:hep-th/0607172].

\bibitem{ks}
  Z.~Komargodski and D.~Shih,
  JHEP {\bf 0904} (2009) 093
  [arXiv:0902.0030 [hep-th]].

\bibitem{ns}
A.~E.~Nelson and N.~Seiberg,
  Nucl.\ Phys.\  B {\bf 416} (1994) 46
  [arXiv:hep-ph/9309299].

\bibitem{iss2}
K.~A.~Intriligator, N.~Seiberg and D.~Shih,
  JHEP {\bf 0707} (2007) 017
  [arXiv:hep-th/0703281].

\bibitem{cw}
S.~R.~Coleman and E.~J.~Weinberg,
  Phys.\ Rev.\  D {\bf 7} (1973) 1888.

\bibitem{grisaru}
 M.~T.~Grisaru, M.~Rocek and R.~von Unge,
  Phys.\ Lett.\  B {\bf 383} (1996) 415
  [arXiv:hep-th/9605149].


\end{thebibliography}
\end{document}